\documentclass[utf8]{frontiersinFPHY_FAMS} 

\setcitestyle{square} 
\usepackage{url,hyperref,lineno,microtype,subcaption}
\usepackage[onehalfspacing]{setspace}
\usepackage{bm}
\usepackage{braket}
\usepackage{amssymb}
\usepackage{graphicx}
\usepackage{dcolumn}
\usepackage{amsmath}
\usepackage{amsfonts}
\usepackage{xcolor}
\usepackage{hyperref}

\def\keyFont{\fontsize{8}{11}\helveticabold }
\def\firstAuthorLast{M.C. Atkinson} 
\def\Authors{M.C. Atkinson\,$^{1,*}$, W.H. Dickhoff\,$^{2}$}
\def\Address{$^{1}$Lawrence Livermore National Laboratory, Livermore, CA, USA\\
$^{2}$Department of Physics, Washington University in St. Louis, St. Louis, MO, USA}
\def\corrAuthor{M.C. Atkinson}

\def\corrEmail{mackenzie.c.atkinson@gmail.com}

\begin{document}
\onecolumn
\firstpage{1}

\title[Neutron skins: A perspective from the DOM]{Neutron skins: A perspective from dispersive optical models} 

\author[\firstAuthorLast ]{\Authors} 
\address{} 
\correspondance{} 

\extraAuth{}

\maketitle

\
\begin{abstract}

An overview is presented of neutron skin predictions obtained using an empirical nonlocal dispersive optical model (DOM). The DOM links both scattering and bound-state experimental data through a subtracted dispersion relation which allows for fully-consistent, data-informed predictions for nuclei where such data exists. Large skins were predicted for both ${}^{48}$Ca ($R^{48}_\textrm{skin}=0.25 \pm 0.023$ fm in 2017) and 
${}^{208}$Pb ($R^{208}_\textrm{skin}=0.25 \pm 0.05$ fm in 2020). While the DOM prediction in $^{208}$Pb is within 1$\sigma$ of the subsequent PREX-2 measurement, the DOM prediction in $^{48}$Ca is over 2$\sigma$ larger than the thin neutron skin resulting from CREX. 
From the moment it was revealed, the thin skin in ${}^{48}$Ca has puzzled the nuclear-physics community as no adequate theories simultaneously predict both a large skin in ${}^{208}$Pb and a small skin in ${}^{48}$Ca. 
The DOM is unique in its ability to treat both structure and reaction data on the same footing, providing a unique perspective on this $R_\textrm{skin}$ puzzle. It appears vital that more neutron data be measured in both the scattering and bound-state domain for ${}^{48}$Ca to clarify the situation.

\tiny
 \keyFont{ \section{Keywords:} neutron skin, structure, reactions, optical potential, Green's function} 
\end{abstract}

\section{Introduction}

A fundamental question in nuclear physics is how the constituent neutrons and protons are distributed in the nucleus. In particular, 
for a nucleus which has a substantial excess of neutrons over protons, are the extra neutrons distributed evenly over the nuclear volume 
or is this excess localized in the periphery of the nucleus forming a neutron skin? 
A quantitative measure is provided by the neutron-skin thickness, $R_\textrm{skin}$, defined as the difference between the point neutron and proton root-mean-squared (RMS) radii, \textit{i.e.}, $R_\textrm{skin} = R_{n} - R_{p}$.

The nuclear symmetry energy, which characterizes the variation of the binding energy as a function of neutron-proton asymmetry, opposes the creation of nuclear matter with excesses of either type of nucleon.
The extent of the neutron skin is determined by the relative strengths of the symmetry energy between the central near-saturation 
and peripheral less-dense regions.  Therefore, $R_\textrm{skin}$ is a measure of the density dependence of the symmetry energy around saturation~\cite{Typel01,Furnstahl02,Steiner05,RocaMaza11}. This dependence is very important for determining many nuclear properties, including masses, radii, and the location of drip lines in the chart of nuclides. Its importance extends to astrophysics for understanding supernovae and neutron stars~\cite{Horowitz01,Steiner10}, and to heavy-ion reactions~\cite{li08}.  
 
Given the rich physics packed in this observable, a large number of studies (both experimental and theoretical) have been devoted to determining neutron skins~\cite{Tsang12,Mammei:2024}. While $R_p$ is extracted quite accurately from elastic electron scattering cross sections (through the charge form factor, $F_\textrm{ch}$)~\cite{Angeli:2013} or laser spectroscopy~\cite{Garcia:2016}, most experimental determinations of $R_n$ are model dependent~\cite{Tsang12}. 
The neutron skin can be determined with essentially the same degree of model independence as $F_\textrm{ch}$ through parity-violating electron scattering~\cite{Horowitz98,Mammei:2024}. The parity-violating asymmetries are governed by the weak form factor, $F_W$, which is the Fourier transform of the weak distribution. The weak distribution is predominantly determined by the neutron distribution, owing to the weak charge of the neutron being of order 1 and that of the proton being nearly 0.   
The first parity-violating experiment performed by the PREX collaboration at Jefferson Lab yielded a thick neutron skin of $^{208}$Pb with a rather large uncertainty~\cite{PREX12}. A second experiment, dubbed PREX-2, was later performed resulting in a $^{208}$Pb skin of $R_\textrm{skin}^{208} = 0.283 \pm 0.071$ fm~\cite{prex2:2021}. The following year, the CREX experiment extracted a much smaller skin in $^{48}$Ca of $R_\textrm{skin}^{48} = 0.121 \pm 0.026(\textrm{exp})\pm 0.024(\textrm{model})$ fm~\cite{crex:2022}.
The large difference between the measured neutron skins in $^{48}$Ca and $^{208}$Pb has puzzled the nuclear-physics community since the CREX result was published.

There currently exists no theory that predicts a thick skin in $^{208}$Pb and a thin skin in $^{48}$Ca. All theoretical studies of these nuclei based on a mean-field approach predict a strong, positive correlation between the neutron skins of $^{208}$Pb and $^{48}$Ca, although it has been argued that the large error bars for PREX-2 may not provide a stringent constraint on the isovector part of energy density functionals~\cite{Reinhard:2022}.
Separate \textit{ab initio} approaches exist for both nuclei.
In Ref.~\cite{Hagen:2016} a neutron skin for ${}^{48}$Ca was predicted that is consistent with the CREX experiment,
while the results of Refs.~\cite{Hu:2022,Hu:2024} exhibit mild tension with the PREX-2 results.
Furthermore, studies of the relation between neutron skins and the nuclear equation of state (EOS) conclude that these skins are tightly correlated with the slope of the symmetry energy, $L$, meaning that the EOS derived from the thin $R_\textrm{skin}$ measured in $^{48}$Ca is incompatible with the EOS derived from the thick $R_\textrm{skin}$ measured in $^{208}$Pb. Through this relation to the nuclear EOS, these differing neutron skin measurements even lead to tensions in exotic astrophysical systems such as neutron stars~\cite{Piekarewicz2024}. More specifically, mass-radius curves predicted from the two different $R_\textrm{skin}$-derived EOS are incompatible with each other and even with observations. 

In this article, we review an alternative theoretical method to predict $R_\textrm{skin}$ in $^{48}$Ca and $^{208}$Pb. 
We employed a dispersive optical model (DOM) analysis of bound and scattering data to constrain the nucleon self-energies, $\Sigma_{\ell j}$, of $^{48}$Ca and $^{208}$Pb.
The self-energy acts as a complex and phenomenological nonlocal potential that unites the nuclear structure and reaction domains~\cite{Mahaux91,Mahzoon:2014,Dickhoff:2017} by leveraging Green's function theory.
 The DOM was originally developed by Mahaux and Sartor~\cite{Mahaux91}, employing local real and imaginary potentials connected through dispersion relations. However, only
with the introduction of nonlocality can realistic self-energies be obtained~\cite{Mahzoon:2014,Dickhoff:2017}. 
The Dyson equation then determines the single-particle propagator, or Green's function, $G_{\ell j}(r,r';E)$, from which bound-state and scattering observables can be deduced. In particular, the particle number and density distributions of the nucleons can be inferred, thus enabling the investigation of neutron skins.
The DOM treats both structure and reaction data on the same footing, unlike mean-field or \textit{ab initio} approaches applied to these systems, providing a unique perspective on the $R_\textrm{skin}$ puzzle revealed by experiments at Jefferson Lab.

The underlying Green's function ingredients of the single-particle propagator are presented in Sec.~\ref{sec:prop} while the DOM framework is introduced in Sec.~\ref{sec:DOM}.
The DOM description of relevant experimental data for ${}^{48}$Ca and ${}^{208}$Pb are presented in Sec.~\ref{sec:fit}.
A discussion of the neutron skin results for these nuclei is given in Sec.~\ref{sec:skin}.
Conclusions and some outlook are presented in Sec.~\ref{sec:conclusions}.

\section{Theory}
\label{sec:theory}
This section is organized to provide brief introductions into the underlying
theory of the DOM.

   \subsection{Single-particle propagator}
\label{sec:prop}
   The single-particle propagator describes the probability amplitude for adding (removing) a particle in state $\alpha$ at one time to the ground state and propagating on top of that state until a later time when it is removed (added) in state $\beta$~\cite{Exposed!}.  In addition to the conserved orbital and
  total angular momentum ($\ell$ and $j$, respectively), the labels $\alpha$ and
  $\beta$ in Eq.~\eqref{eq:green} refer to a suitably chosen single-particle basis. 
  We employed a coordinate-space basis in our original $^{48}$Ca calculation in Ref.~\cite{Mahzoon:2017} but have since updated to using a Lagrange basis~\cite{Baye:2010} in all subsequent calculations (including that of $^{208}$Pb from Ref.~\cite{Atkinson20}).
 It is convenient to work with the Fourier-transformed propagator in the energy domain, 
 \begin{align}
    G_{\ell j}(\alpha,\beta;E)  &= \bra{\Psi_0^A}a_{\alpha \ell j}
    \frac{1}{E-(\hat{H}-E_0^A)+i\eta} a_{\beta \ell j}^\dagger\ket{\Psi_0^A}
    \nonumber \\ 
    + \bra{\Psi_0^A}&a_{\beta \ell j}^\dagger\frac{1}{E-(E_0^A-\hat{H})-i\eta}
    a_{\alpha \ell j}\ket{\Psi_0^A},
    \label{eq:green}
 \end{align}
 with $E^A_0$ representing the energy of the nondegenerate ground state $\ket{\Psi^A_0}$.
 Many interactions can occur between the addition and removal of the particle (or \textit{vice versa}), all of which need to be considered to calculate the propagator. 
 No assumptions about the detailed form of the Hamiltonian $\hat{H}$ need be made for the present discussion, but it will be assumed that a meaningful Hamiltonian exists that contains two-body and three-body contributions.
 Application of perturbation theory then leads to the Dyson equation~\cite{Exposed!} given by
 \begin{equation}
    G_{\ell j}(\alpha,\beta;E) = G_{\ell}^{(0)}(\alpha,\beta;E)  +
    \sum_{\gamma,\delta}G_{\ell}^{(0)}(\alpha,\gamma;E)\Sigma_{\ell
    j}^*(\gamma,\delta;E)G_{\ell j}(\delta,\beta;E) ,
    \label{eq:dyson}
 \end{equation}
 where $G^{(0)}_{\ell}(\alpha,\beta;E)$ corresponds to the unperturbed propagator (the propagator derived from the unperturbed Hamiltonian, $H_0$, which in the DOM corresponds to the kinetic energy)
 and $\Sigma_{\ell j}^*(\gamma,\delta;E)$ is the irreducible self-energy~\cite{Exposed!}. 
 The hole spectral density for energies below $\varepsilon_F$ is obtained from 
 \begin{equation}
    S^h_{\ell j}(\alpha,\beta;E) = \frac{1}{\pi}\textrm{Im}\ G_{\ell j}(\alpha,\beta;E), 
    \label{eq:spec}
 \end{equation}
 where the $h$ superscript signifies it is the hole spectral amplitude. For brevity, we drop this superscript for the rest of this review.
 The diagonal element of Eq.~\eqref{eq:spec} is known as the (hole) spectral function identifying the probability density for the removal of a single-particle state with quantum numbers $\alpha \ell j$ at
 energy $E$. The single-particle density distribution can be calculated from the hole spectral function in the following way, 
 \begin{equation}
    \label{eq:density}
    \rho_{\ell j}^{(p,n)}(r) = \sum_{\ell j} (2j+1) \int_{-\infty}^{\varepsilon_F}dE\ S^{(p,n)}_{\ell j}(r,r;E),
 \end{equation}
 where the $(p,n)$ superscript refers to protons or neutrons, and $\varepsilon_F = \frac{1}{2}(E^{A+1}_0 - E^{A-1}_0)$ is the average Fermi energy which separates the particle and hole domains~\cite{Exposed!}.
 The number of protons and neutrons ($Z,N$) is calculated by integrating $\rho_{\ell j}^{(p,n)}(r)$ over all space. In addition to particle number, the total binding energy can be calculated from the hole spectral function using the Migdal-Galitski sum rule~\cite{Exposed!}, 
 \begin{equation}
    E_0^{N,Z} = \frac{1}{2}\sum_{\alpha\beta}\int_0^{\varepsilon_F}dE\left[\braket{\alpha|\hat{T}|\beta}S^h(\alpha,\beta;E)
     + \delta_{\alpha\beta}ES^h(\alpha,\alpha;E)\vphantom{\braket{\alpha|\hat{T}|\beta}S^h(\alpha,\beta;E)}\right].
    \label{eq:energy_sumrule}
 \end{equation}
This expression assumes that the dominant contribution involves the two-nucleon interaction~\cite{Atkinson2020B,Atkinson2021} and the $\ell j$ labels have been subsumed in $\alpha$ and $\beta$.

To visualize the spectral function of Eq.~\eqref{eq:spec}, it is useful to sum (or integrate) over the basis variables, $\alpha$, so that only an energy-dependence remains, $S_{\ell j}(E)$.
 The spectral strength $S_{\ell j}(E)$ is the contribution at energy $E$ to the occupation from all orbitals with angular momentum $\ell j$. It reveals that the strength for a shell can be fragmented, rather than isolated
 at the independent-particle model (IPM) energy levels.  
 \begin{figure*}[t]
    \begin{center}
       \includegraphics[scale=1.0]{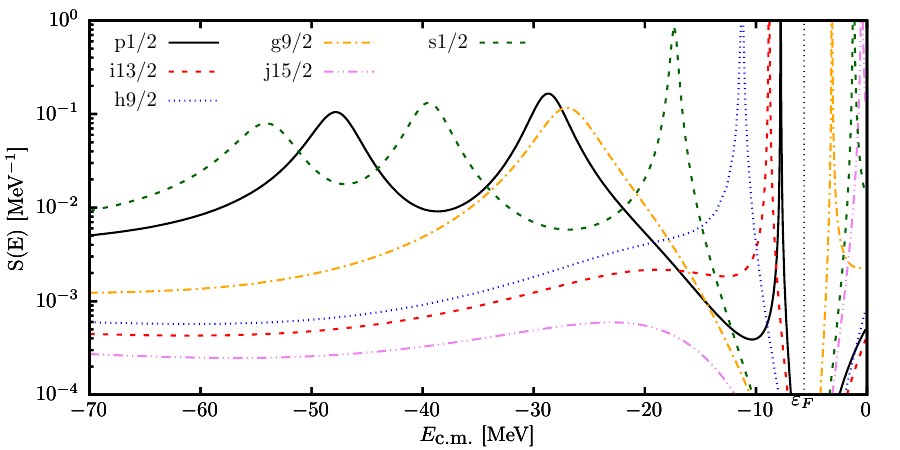}
    \end{center}
    \caption{\label{fig:spectral_n}Neutron spectral functions of a representative set of $\ell j$ shells in $^{208}$Pb. The particle states are differentiated from the hole states by the vertical dashed line which corresponds to the location of the Fermi energy. Figure adapted from Ref.~\cite{Atkinson:2020}}

 \end{figure*}
 Figure~\ref{fig:spectral_n} shows the spectral strength for a representative set of neutron shells in $^{208}$Pb that would be considered bound and fully occupied in the IPM. 
 The  location of the peaks in Fig.~\ref{fig:spectral_n} correspond to the energies of discrete bound states with one nucleon removed. For example, the s$1/2$ spectral function in Fig.~\ref{fig:spectral_n} has four peaks, three below
 $\varepsilon_F$ corresponding to the 0s$1/2$, 1s$1/2$, and 2s$1/2$ quasihole states, and one above $\varepsilon_F$ corresponding to the 3s$1/2$ quasiparticle state.  
 The quasihole wave functions of these bound states can be obtained by transforming the Dyson equation into a nonlocal Schr\"{o}dinger-like equation by disregarding the imaginary part of $\Sigma^*(\alpha,\beta;E)$,
 \begin{align}
    \sum_\gamma\bra{\alpha}T_{\ell} + \textrm{Re}\ \Sigma^*_{\ell j}(\varepsilon_{\ell j}^n)\ket{\gamma}\psi_{\ell j}^n(\gamma) = \varepsilon_{\ell j}^n\psi_{\ell j}^n(\alpha),
    \label{eq:schrodinger}
 \end{align}
 where $\bra{\alpha}T_\ell\ket{\gamma}$ is the kinetic-energy matrix element, including the centrifugal term.
 The wave function, $\psi_{\ell j}^n(\alpha)$, is the overlap between the $A$ and $A-1$ systems and the corresponding energy, $\varepsilon_{\ell j}^n$, is the energy required to remove a nucleon with the particular quantum numbers $n\ell j$, 
 \begin{equation}
    \psi^n_{\ell j}(\alpha) = \bra{\Psi_n^{A-1}}a_{\alpha \ell j}\ket{\Psi_0^A}, \qquad \varepsilon_{\ell j}^n = E_0^A - E_n^{A-1}.
    \label{eq:wavefunction}
 \end{equation}

 When solutions to Eq.~\eqref{eq:schrodinger} are found near the Fermi energy where there is naturally no imaginary part of the self-energy, the normalization of the quasihole is well-defined as the spectroscopic factor, 
 \begin{equation}
    \mathcal{Z}^n_{\ell j} = \bigg(1 - \frac{\partial\Sigma_{\ell j}^*(\alpha_{qh},\alpha_{qh};E)}{\partial E}\bigg|_{\varepsilon_{\ell j}^n}\bigg)^{-1},
    \label{eq:sf}
 \end{equation}
 where $\alpha_{qh}$ corresponds to the quasihole state that solves Eq.~\eqref{eq:schrodinger}. The quasihole peaks in Fig.~\ref{fig:spectral_n} get narrower as the levels approach $\varepsilon_F$, which is a consequence of the imaginary part of the irreducible self-energy decreasing when
 approaching $\varepsilon_F$. In fact, the last mostly occupied neutron level in Fig.~\ref{fig:spectral_n} (2p$1/2$) has a spectral function that is essentially a delta function peaked at its energy level, where
 the imaginary part of the self-energy vanishes.  For these orbitals, the strength of the spectral function at the peak corresponds to the spectroscopic factor in Eq.~\eqref{eq:sf}. The spectroscopic factor can be probed using the exclusive $(e,e'p)$ reaction which will be discussed in Sec.~\ref{sec:dom-predict} (see also Refs.~\cite{Atkinson:2018,Atkinson:2019}). 

 \subsection{Dispersive optical model}
 \label{sec:DOM}

 The Dyson equation, Eq.~\eqref{eq:dyson}, simplifies the complicated task of calculating $G(\alpha,\beta;E)$ from Eq.~\eqref{eq:green} to finding and inverting a suitable $\Sigma^*(\alpha,\beta;E)$ (suppressing the $\ell j$ labels).
 It was recognized long ago that $\Sigma^*(\alpha,\beta;E)$ represents the potential 
 that describes elastic-scattering observables~\cite{Bell59}. 
 The link with the potential at negative energy is then provided by the Green's function framework as was realized by Mahaux and Sartor who introduced the DOM as reviewed in Ref.~\cite{Mahaux91}. 
 The analytic structure of the nucleon self-energy allows one 
 to apply the dispersion relation, which relates the real part of the self-energy at a given energy to a dispersion integral of its imaginary part over all energies.
 The energy-independent correlated Hartree-Fock (HF) contribution~\cite{Exposed!} is removed by employing a subtracted dispersion relation with the Fermi energy used as the subtraction point~\cite{Mahaux91}.
 The subtracted form has the further advantage that the emphasis is placed on energies closer to the Fermi energy for which more experimental data are available.
 The real part of the self-energy at the Fermi energy is then still referred to as the HF term, and is sufficiently attractive to bind the relevant levels at about the correct energies.
 In practice, the imaginary part is assumed to extend to the Fermi energy on both sides while being very small in its  vicinity.
 The subtracted form of the dispersion relation employed in this work is given by
 \begin{align}
    \textrm{Re}\ \Sigma^*(\alpha,\beta;E) &= \textrm{Re}\
    \Sigma^*(\alpha,\beta;\varepsilon_F) \label{eq:dispersion} \\ -
    \mathcal{P}\int_{\varepsilon_F}^{\infty} \!\! \frac{dE'}{\pi}&\textrm{Im}\
    \Sigma^*(\alpha,\beta;E')\left[\frac{1}{E-E'}-\frac{1}{\varepsilon_F-E'}\right] \nonumber
    \\ + \mathcal{P} \! \int_{-\infty}^{\varepsilon_F} \!\!
    \frac{dE'}{\pi}&\textrm{Im}\
    \Sigma^*(\alpha,\beta;E')\left[\frac{1}{E-E'}-\frac{1}{\varepsilon_F-E'}\right],
    \nonumber      
 \end{align}
 where $\mathcal{P}$ is the principal value. 
 The static term, $\textrm{Re}\Sigma^*(\alpha,\beta;\varepsilon_F)$, is denoted by $\Sigma_{\text{HF}}$ from here on. 
 Equation~\eqref{eq:dispersion} constrains the real part of $\Sigma^*(\alpha,\beta;E)$ by empirical information of its HF and imaginary parts which are closely tied to experimental data. 
 Initially, standard functional forms for these terms were introduced by Mahaux and Sartor who also cast the DOM potential in a local form by a standard transformation which turns a nonlocal static HF potential into an energy-dependent local potential~\cite{Perey:1962}.
 Such an analysis was extended in Refs.~\cite{Charity06,Charity:2007} to a sequence of Ca isotopes and in Ref.~\cite{Mueller:2011} to semi-closed-shell nuclei heavier than Ca.
 The transformation to the exclusive use of local potentials precludes a proper calculation of nucleon particle number and expectation values of the one-body operators, like the charge density in the ground state (see Eq.~\eqref{eq:density}). 
 This obstacle was eliminated in Ref.~\cite{Dickhoff:2010}, but it was shown that the introduction of nonlocality in the imaginary part was still necessary in order to accurately account for particle number and the charge density~\cite{Mahzoon:2014}.
 Theoretical work provided further support for this introduction of a nonlocal representation of the imaginary part of the self-energy~\cite{Waldecker:2011,Dussan:2011}.
 A review detailing these developments was published in Ref.~\cite{Dickhoff:2017}.

 \subsubsection{Functional Form of DOM Self-Energy}

 We employ a nonlocal representation of the self-energy following
 Ref.~\cite{Mahzoon:2014} where $\Sigma_{\text{HF}}(\bm{r},\bm{r'})$ and
 $\textrm{Im}\ \Sigma(\bm{r},\bm{r'};E)$ are parametrized and the energy-dependence of the real part, $\textrm{Re}\ \Sigma(\bm{r},\bm{r'};E)$, is generated from the dispersion relation in Eq.~\eqref{eq:dispersion}. The HF term consists of a volume term, spin-orbit term,  and a wine-bottle-shaped term~\cite{Brida11},
\begin{align}
\Sigma_{HF}(\bm{r},\bm{r}') = V_\textrm{vol}(\bm{r},\bm{r}') + V_\textrm{so}(\bm{r},\bm{r}') + V_\textrm{wb}(\bm{r},\bm{r}') + \delta(\bm{r}-\bm{r}') V_C(r),
\end{align}
where the Coulomb potential, $V_C(r)$, is also included.
The radial part of our potentials takes the following form, 
\begin{equation}
   V_\textrm{vol}\left( \bm{r},\bm{r}' \right) =  V^\textrm{vol}
   \,f \left ( \tilde{r},r^\textrm{HF}_{(p,n)},a^\textrm{HF} \right ) 
   H \left( \bm{s};\beta^\textrm{HF} \right),
   \label{eq:HFvol} 
\end{equation}
where $V^\textrm{vol}$ is a parameter that determines the depth of the potential and $r^\textrm{HF}_{(p,n)}$, $a^{\textrm{HF}}$, and $\beta^\textrm{HF}$ are parameters that control the shape of the Woods-Saxon form factor $f$ and Perey-Buck-shaped~\cite{Perey:1962} nonlocality $H$, 
\begin{equation}
   f(r,r_{i},a_{i})=\left[1+\exp \left({\frac{r-r_{i}A^{1/3}}{a_{i}}%
   }\right)\right]^{-1} \hspace{1cm}
   H \left( \bm{s}; \beta \right) = \exp \left( - \bm{s}^2 / \beta^2 \right)/ (\pi^{3/2} \beta^3),
\end{equation}
and
\begin{equation}
   \tilde{r} =\frac{r+r'}{2} \hspace{3cm} \bm{s}=\bm{r}-\bm{r}'.
\label{eq:com_coords}
\end{equation}
Nonlocality is introduced in a similar way for $V_\textrm{wb}(\bm{r},\bm{r}')$ and $V_\textrm{so}(\bm{r},\bm{r}')$; their explicit forms can be found in Ref.~\cite{Atkinson:2020}.
The imaginary self-energy consists of volume, surface, and spin-orbit terms, 
\begin{align}
   \textrm{Im}\Sigma(\bm{r},\bm{r}';E) &= 
   -W^{vol}_{0\pm}(E) f\left(\tilde{r};r^{vol}_{\pm};a^{vol}_{\pm}\right)H \left( \bm{s}; \beta^{vol}\right) \nonumber \\
   &+ 4a^{sur}_\pm W^{sur}_{\pm}\left( E\right)H \left( \bm{s}; \beta^{sur}\right)
   \frac{d}{d \tilde{r} }f(\tilde{r},r^{sur}_{\pm},a^{sur}_{\pm})  + \textrm{Im}\Sigma_{so}(\bm{r},\bm{r}';E),
   \label{eq:imnl}
\end{align}
where $W^{vol}_{0\pm}(E)$ and $W^{sur}_{\pm}\left( E\right)$ are energy-dependent depths of the volume and surface potentials, respectively, and the $\pm$ subscript indicates there are different forms used above and below the Fermi energy (see Ref~\cite{Atkinson:2020} for exact forms). When considering asymmetric nuclei, such as $^{48}$Ca and $^{208}$Pb, additional terms proportional to the asymmetry, $\alpha = \frac{A-Z}{A}$, are added to $\Sigma_\textrm{HF}(\bm{r},\bm{r}')$ and $\textrm{Im}\Sigma(\bm{r},\bm{r}';E)$ for a Lane-like representation~\cite{Lane62}.
These asymmetric terms introduce additional parameters describing both their radial shape and energy-dependent depths~\cite{Atkinson:2020}. See Refs.~\cite{Atkinson:2020,Atkinson:2019} for the full list of parameters used in $^{48}$Ca and $^{208}$Pb.

 As mentioned previously, it was customary in the past to replace nonlocal potentials by local, energy-dependent potentials~\cite{Mahaux91,Perey:1962,Fiedeldey:1966,Exposed!}. The introduction of an energy dependence alters the dispersive
 correction from Eq.~\eqref{eq:dispersion} and distorts the normalization, leading to incorrect spectral functions and related quantities~\cite{Dickhoff:2010}. Thus, a nonlocal implementation permits the self-energy to accurately 
 reproduce important observables such charge density, particle number, and ground-state binding energy.

 \subsection{DOM fits of $^{208}$Pb and $^{48}$Ca}
 \label{sec:fit}

 To use the DOM self-energy for predictions, the parameters of the self-energy are constrained through weighted $\chi^2$ minimization (using the Powell method~\cite{Numerical}) by measurements of elastic differential cross sections ($\frac{d\sigma}{d\Omega}$), analyzing powers ($A_\theta$), reaction cross sections ($\sigma_\textrm{react}$), total cross sections ($\sigma_\textrm{tot}$), charge density ($\rho_{\textrm{ch}}$), energy levels ($\varepsilon_{n \ell j}$), particle number, and the root-mean-square charge radius ($R_{\text{ch}}$).
The angular-dependence of $\Sigma(\bm{r},\bm{r}';E)$ is represented in a partial-wave basis, and the radial component is represented in a Lagrange basis using Legendre and Laguerre polynomials for scattering and bound states, respectively.
 The bound states are found by diagonalizing the Hamiltonian in
 Eq.~\eqref{eq:schrodinger}, the propagator is found by inverting the Dyson
 equation, Eq.~\eqref{eq:dyson}, 
 while all scattering calculations are done in the framework of $R$-matrix theory~\cite{Baye:2010}. 
 While it has been suggested in Refs.~\cite{Khoa:2007,Loc:2014,Pawel17} that charge-exchange reactions to
isobaric analogue states could further constrain the isovector
potential, charge-exchange data were not included in the fits reviewed in this article. 
Reasonable cross sections are obtained with our DOM potential,
suggesting that these data, while important, are not sufficient to alter the
conclusions of our work significantly. This may be due to the use of nonlocal
potentials as opposed to the local ones used in Refs.~\cite{Loc:2014,Khoa:2007} based on Ref.~\cite{Koning:2003}.


When constraining the $^{48}$Ca self-energy, the isoscalar part is largely determined by the nearby $N=Z$ $^{40}$Ca nucleus. Therefore, using our $^{40}$Ca parametrization from Ref.~\cite{Atkinson:2018} as a starting point, we only needed to fit the asymmetric parameters of the $^{48}$Ca potential~\cite{Mahzoon:2017,Atkinson:2020}. This resulted in a $^{48}$Ca self-energy that closely reproduced all training data~\cite{Mahzoon:2017}. In the case of $^{208}$Pb, there is not a nearby nucleus with $N=Z$, therefore we started from the $^{48}$Ca parameters of Ref.~\cite{Atkinson:2019} and varied both the isoscalar and isovector parameters to reproduce experimental data. To illustrate how well this method works, we show the result of the $^{208}$Pb fit below.



\begin{figure}[h]
    \begin{center}
          \includegraphics[width=0.49\linewidth]{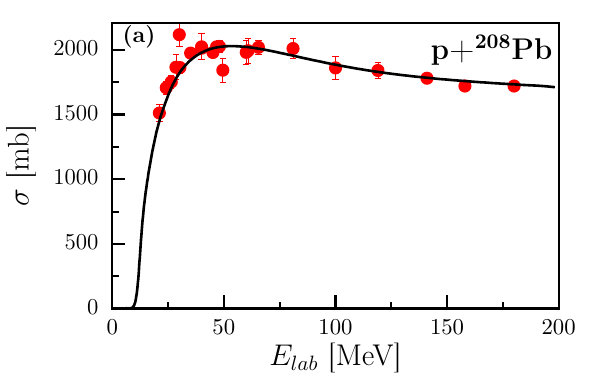}
          \includegraphics[width=0.49\linewidth]{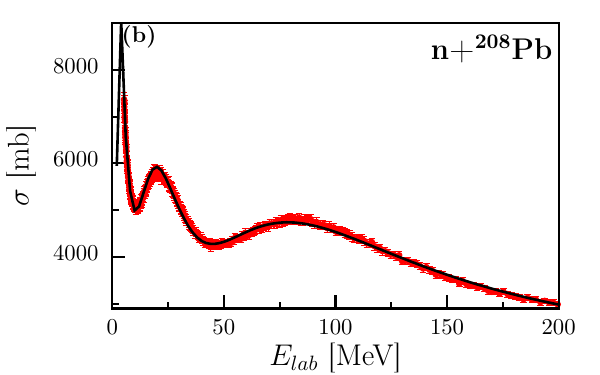}
       \end{center}
    \caption{(a) Proton reaction
       cross section in $^{208}$Pb. The solid line is generated from the DOM self-energy while the filled circles are from experiment. (b) Neutron total cross section in $^{208}$Pb. The solid line is generated from the DOM
       self-energy for $^{208}$Pb while the filled circles are from experiment.  See Ref.~\cite{Mueller:2011} for the experimental data. Figure adapted from Ref.~\cite{Atkinson:2020}.
     \label{fig:react_pb208}}
   
 \end{figure}

\begin{figure}[h]
    \begin{center}
          \includegraphics[scale=0.65]{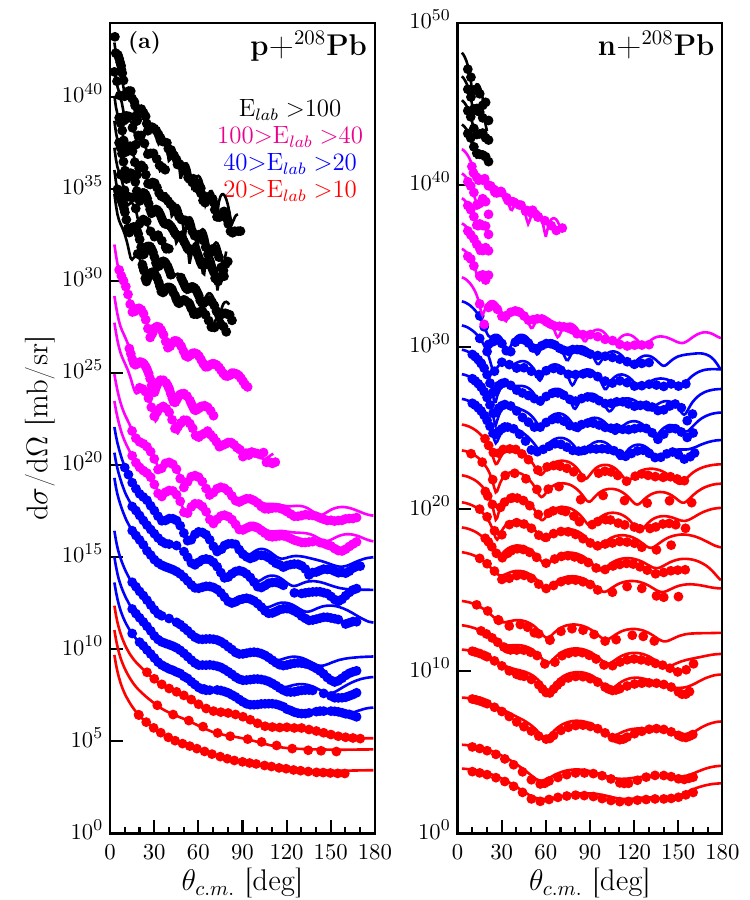}
          \includegraphics[scale=0.65]{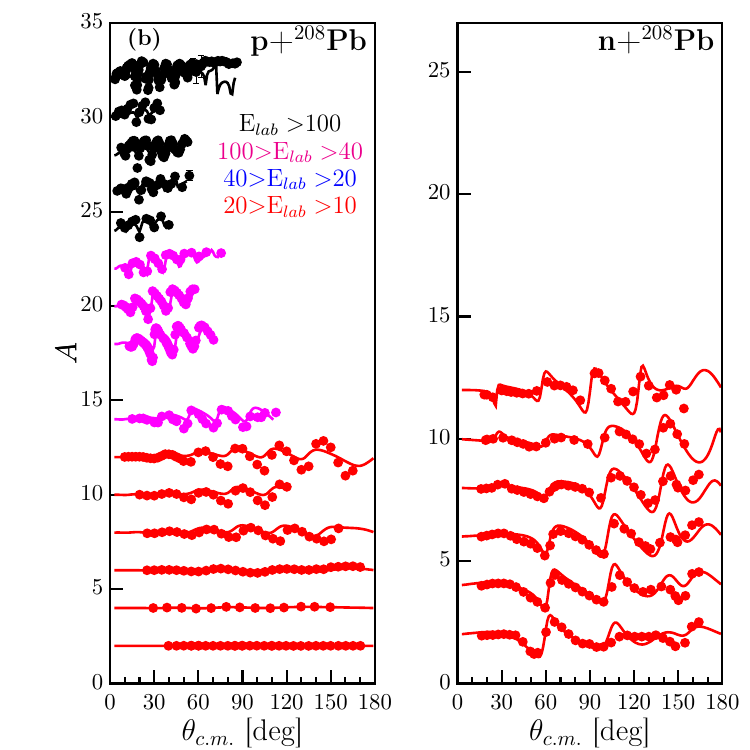}
       \end{center}
    \caption[Calculated and experimental proton and neutron elastic-scattering
       angular distributions of the differential cross section
    $\frac{d\sigma}{d\Omega}$ for $^{208}$Pb]{(a) Calculated and experimental proton and neutron
       elastic-scattering angular distributions of the differential cross section
       $\frac{d\sigma}{d\Omega}$ for $^{208}$Pb ranging from 10 MeV to 200 MeV. The data at each energy is offset by factors of
       ten to help visualize all of the data at once.  
       (b) Results for proton
       and neutron analyzing power generated from the DOM self-energy for
       $^{208}$Pb compared with experimental data ranging from 10 MeV to 200 MeV. References to the data are given
       in Ref.~\cite{Mueller:2011}. Figure adapted from Ref.~\cite{Atkinson:2020}.
    \label{fig:elastic_pb208}
    }
 \end{figure}

 Proton reaction cross sections together with the DOM result are displayed in panel (a) of Fig.~\ref{fig:react_pb208}.
 The neutron total cross section is shown in panel (b) of Fig.~\ref{fig:react_pb208}.  
 Both aggregate cross sections play an important role in determining volume integrals of the imaginary part of the self-energy, thereby providing strong constraints on the depletion of IPM orbits.
 The elastic
 differential cross sections of proton and neutrons up to 200 MeV are shown in panel (a) of Fig.~\ref{fig:elastic_pb208}. Panel (b) contains the analyzing powers for neutrons and protons which strongly constrain the spin-orbit components of the self-energy.

 \begin{figure}[h]
    \begin{minipage}{\linewidth}
       \makebox[\linewidth]{
          \includegraphics[scale=1.0]{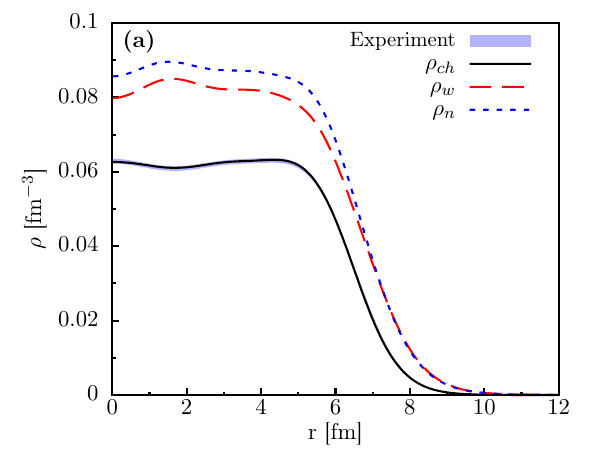}
          \includegraphics[scale=1.0]{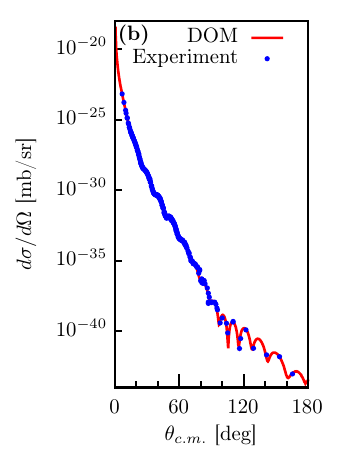}
       }
    \end{minipage}
    \caption[Experimental and fitted $^{208}$Pb charge density.]{ \label{fig:chd_pb208}(a) Experimental
       and fitted $^{208}$Pb charge density. The solid black line is calculated
       using Eq.~\eqref{eq:density} and folding with the proton
       charge distribution while the experimental band represents the 1\%
       error associated with the extracted charge density from elastic
       electron scattering experiments using the sum of Gaussians
       parametrization~\cite{deVries:1987,Sick79}. Also shown is the deduced weak charge distribution, $\rho_w$ (red long-dashed line), and neutron matter distribution, $\rho_n$ (blue short-dashed line).
       (b) Experimental and fitted elastic electron scattering differential cross section in $^{208}$Pb. All available data have been transformed to an electron energy of 502 MeV in the center-of-mass frame~\cite{Frois77}. Figure adapted from Ref.~\cite{Atkinson:2020}.
    }
 \end{figure} 

 The charge density of $^{208}$Pb is shown in panel (a) of Fig.~\ref{fig:chd_pb208}. The
 experimental band is extracted from elastic electron scattering differential
 cross sections~\cite{deVries:1987}. This data set is well reproduced after using the
 DOM charge density from Fig.~\ref{fig:chd_pb208} as the ingredient in a
 relativistic elastic electron scattering code~\cite{salvat:2005}. The corresponding elastic electron 
 scattering cross section is shown in panel (b) of Fig.~\ref{fig:chd_pb208} and compared to experiment with all available data transformed to an electron energy of 502 MeV in the center-of-mass frame~\cite{Frois77}.

\begin{figure}[h]
   \begin{center}
         \includegraphics[scale=0.95]{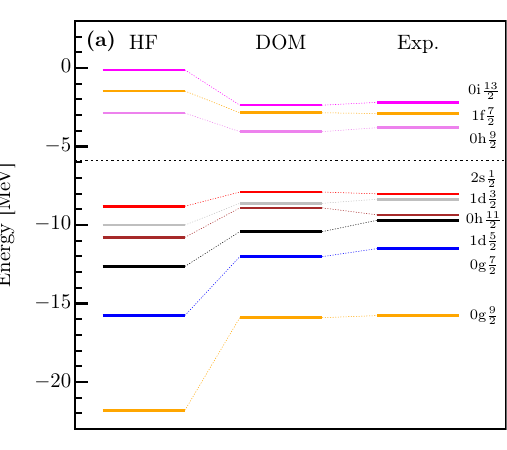}
         \includegraphics[scale=0.95]{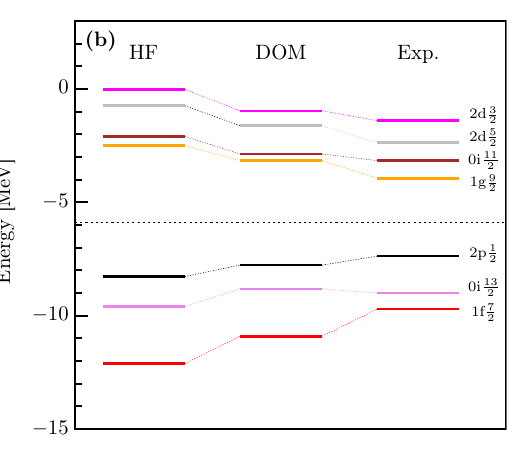}
      \end{center}
   \caption{ \label{fig:levelsP}(a) Proton and (b) neutron energy levels in $^{208}$Pb. The energies on the left are calculated using only the static part of the DOM self-energy, corresponding to a Hartree-Fock calculation. The middle energies are those calculated using the full DOM self-energy. The energies on the right correspond to the experimental values. The change from the left energies to the middle energies is the result of including the dynamic part of the self-energy. Figure adapted from Ref.~\cite{Atkinson:2020}.}
\end{figure}

In Fig.~\ref{fig:levelsP}, single-particle levels calculated using Eq.~\eqref{eq:schrodinger} are compared to the experimental values for protons and neutrons in panels (a) and (b), respectively. 
The middle column consists of levels calculated using the full DOM and the right column contains the experimental levels. The first column of the figures represents a calculation using only the static part of
the self-energy, corresponding to the Hartree-Fock (mean-field) contribution. It is clear from these level diagrams that the mean-field overestimates the particle-hole gap (see also Ref.~\cite{Bender:2003}). The inclusion of the dynamic part of
the self-energy is necessary to reduce this gap and properly describe the energy levels~\cite{Mahaux91}. Furthermore, the effect of including the dynamic part of the self-energy on the proton levels is stronger than the
effect on the neutron levels. This suggests that protons deviate more from the IPM than neutrons in $^{208}$Pb.

The number of neutrons and protons in the DOM fit of $^{208}$Pb, calculated by integrating Eq.~\eqref{eq:density} using shells up to $\ell \le 20$, is shown in Table~\ref{table:sumrules}.  As there are 82 protons and 126 neutrons in $^{208}$Pb, the reported values are accurate to within a fraction of a percent.  The binding energy of $^{208}$Pb was fit to the experimental value using Eq.~\eqref{eq:energy_sumrule}. 
As there is no way at present to assess the contribution of three-body interactions to the ground-state energy, we employ the present approximation which applies when only two-body interactions occur in the Hamiltonian, to ensure that enough spectral strength occurs at negative energy which has implications for the presence of high-momentum components. Also shown in Table~\ref{table:sumrules} is $R^{208}_\textrm{ch}$ calculated as the RMS radius of the charge density displayed in Fig.~\ref{fig:chd_pb208}.  

\begin{table}[b]
   \begin{minipage}{\linewidth}
      \makebox[\linewidth]{
         {\renewcommand{\arraystretch}{1.15}
         \begin{tabular}{c  c c  c c} 
            \hline
            \hline
            & N & Z & $E_0^A/A$ [MeV] & $R_\textrm{ch}$ [fm]\\
            \hline
            DOM & 126.2 & 82.08 & -7.82 & 5.48 \\
            Expt. & 126 & 82 & -7.87 & 5.50 \\
            \hline
            \hline
         \end{tabular}
      }
   }
\end{minipage}
\caption{ \label{table:sumrules}Comparison of the calculated DOM particle numbers and binding energy of $^{208}$Pb and the corresponding experimental values. The experimental binding energy was taken from Ref.~\cite{AME2020}. The experimental charge radius is from Ref.~\cite{deVries:1987}}
\end{table}
The reproduction of all available experimental data indicates that we have realistic self-energies of $^{208}$Pb and similarly for $^{48}$Ca~\cite{Mahzoon:2017,Atkinson:2019} capable of describing both bound-state and scattering processes. With these self-energies we can therefore make predictions of observables such as the neutron skin. Additionally, a parallel DOM analysis of these and other nuclei was conducted using Markov Chain Monte Carlo (MCMC) to optimize the potential parameters employing the same experimental data and a very similar functional form but with a reduced number of parameters. All observables from this MCMC fit fell within one standard deviation of those presented above~\cite{Pruitt:2020,Pruitt:2020C}.

\subsection{DOM Predictions}
\label{sec:dom-predict}


Spectroscopic factors come directly from the self-energy through Eq.~\eqref{eq:sf}, making the DOM ideal for predicting $(e,e'p)$ cross sections (see the $^{40}$Ca$(e,e'p)^{39}$K analysis in Ref.~\cite{Atkinson:2018}). When we tried to calculate $^{48}$Ca$(e,e'p)^{47}$K using the fit from Ref.~\cite{Mahzoon:2017}, we found that the spectroscopic factors were too large to describe the data. Unlike in $^{40}$Ca, there is a lack of high-energy ($E>100$ MeV) proton reaction cross-section data in $^{48}$Ca. This allowed the fit of Ref.~\cite{Mahzoon:2017} to predict proton reaction cross sections which fell off for higher energies. Consequentially, the proton spectroscopic factors which were too large to describe $^{48}$Ca$(e,e'p)^{47}$K data to the same degree of accuracy achieved for $^{40}$Ca~\cite{Atkinson:2018}. Observing for $E_\textrm{c.m.}> 150$ MeV that $\sigma_\textrm{react}(E)$ is close to constant, we used the ratio of $\sigma_\textrm{react}(E)$ measurements of $^{40}$Ca and $^{48}$Ca at 700 MeV~\cite{Anderson700} to scale the $^{40}$Ca $\sigma_\textrm{react}(E)$ data such that it could be used as a constraint for $^{48}$Ca. Thanks to the dispersion relation, Eq.~\eqref{eq:dispersion}, the increased $\textrm{Im}\Sigma(r,r';E)$ to accommodate a higher reaction cross sections at positive energies pulls strength from below $\varepsilon_F$. This reduced the spectroscopic factors which then allowed for accurate descriptions of $^{48}$Ca$(e,e'p)^{47}$K cross sections~\cite{Atkinson:2019}. This only altered the proton parameters, thus the neutron skin remained unchanged at $R_\textrm{skin} = 0.25$ fm. This demonstrates that once a sufficiently-complete set of data is used, the DOM is capable of making accurate predictions.


The valence spectroscopic factors in $^{208}$Pb are consistent with the observations of Ref.~\cite{Lichtenstadt79} and the interpretation of Ref.~\cite{Vijay84}.
The past extraction of spectroscopic factors using the $(e,e'p)$ reaction yielded a value around 0.65 for the valence 
$\mathrm{2s_{1/2}}$ orbit~\cite{Ingo91} based on the results of Ref.~\cite{Quint86,Quint87}.
While the use of nonlocal optical potentials may slightly increase this value as shown in Ref.~\cite{Atkinson:2018}, it may be concluded that the value of 0.69 obtained from the DOM analysis is consistent with the past result.
Nikhef data obtained in a large missing energy and momentum domain~\cite{Batenburg01} can now be consistently analyzed employing the complete DOM spectral functions.

\begin{figure}[t]
   \begin{center}
      {
         \includegraphics{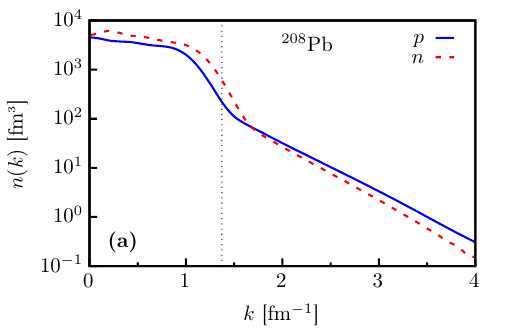}
         \includegraphics{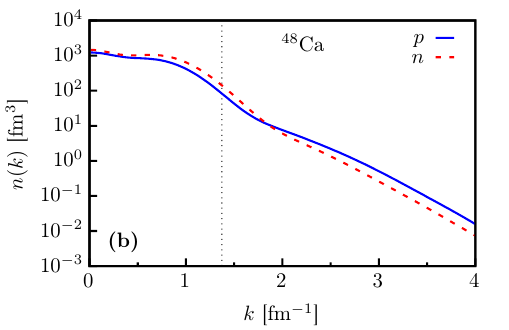}
      }
   \end{center}
   \caption[Comparison of calculated DOM momentum distribution of protons and
   neutrons in $^{208}$Pb.]
   { \label{fig:kdist}
      Comparison of calculated DOM momentum distributions of protons (solid blue line) and neutrons (dashed red line). The vertical dotted line marks the location of $k_F$. (a) Momentum distributions in $^{208}$Pb. (b) Momentum distributions in $^{48}$Ca. Figure adapted from Refs.~\cite{Atkinson:2019,Atkinson:2020}.
   }
\end{figure}

Correlations can also be studied through the momentum distribution, $n(k)$, which represents the diagonal of the double Fourier-transform of the single-particle density matrix. 
The calculated DOM momentum distributions of $^{48}$Ca and $^{208}$Pb are shown in Fig.~\ref{fig:kdist}. The high-momentum tail of $n(k)$ arises from short-range correlations (SRC), which is another
manifestation of many-body correlations beyond the IPM description of the nucleus~\cite{Hen:2017}. This high-momentum content can be quantified by integrating the momentum distribution above the Fermi momentum.
Using $k_F=270$ MeV/c, 13.4\% of protons and 10.7\% of neutrons have momenta greater than $k_F$ in $^{208}$Pb whereas $^{48}$Ca has 14.6\% high-$k$ protons and 12.6\% high-$k$ neutrons.  These numbers are in qualitative agreement with what is observed in the high-momentum knockout experiments done by the CLAS
collaboration at Jefferson Lab~\cite{CLAS:2006}.  Furthermore, the fraction of high-momentum protons is larger than the fraction of high-momentum neutrons. 
These features were predicted by \textit{ab initio} calculations of asymmetric nuclear matter reported in Refs.~\cite{Frick:2005,Rios:2009,Rios:2014} which demonstrated unambiguously that the inclusion of the nucleon-nucleon tensor force, when constrained by nucleon-nucleon scattering data, is responsible for making protons more correlated with increasing nucleon asymmetry at normal density.
This supports the $np$-dominance picture in which
the dominant contribution to SRC pairs comes from $np$ SRC pairs which arise from the tensor force in the nucleon-nucleon interaction~\cite{Duer:2018,Wiringa:2014}. Due to the neutron excess in $^{208}$Pb and $^{48}$Ca,
there are more neutrons available to make $np$ SRC pairs which leads to an increase in the fraction of high-momentum protons. 

In the DOM, this high-momentum content is determined by how much strength exists in the hole spectral function at large, negative energies. The hole spectral function is constrained in the fit by the particle
number, binding energy, and charge density. While the particle number and charge density can only constrain the total strength of the hole spectral function, the binding energy constrains how the strength of
the spectral function is distributed in energy. This arises from the energy-weighted integral in Eq.~\eqref{eq:energy_sumrule}, which will push strength of the spectral function to more-negative
energies in order to achieve more binding. This, in turn, alters the momentum distribution, thus partially constraining the high-momentum content.
It should be noted that the DOM does not exhibit the characteristic energy dependence of high-momentum strength distributions~\cite{Rohe2004} as reported in Ref.~\cite{Mahzoon:2014}.
Such a dependence is more difficult to implement as it requires abandoning the factorization of spatial and energy dependence of the DOM self-energy [see Eq.~(\ref{eq:imnl})].

\section{Neutron Skin}
\label{sec:skin}

\begin{figure}[t]
   \begin{minipage}{\linewidth}
      \makebox[\linewidth]{
         \includegraphics[scale=1.4]{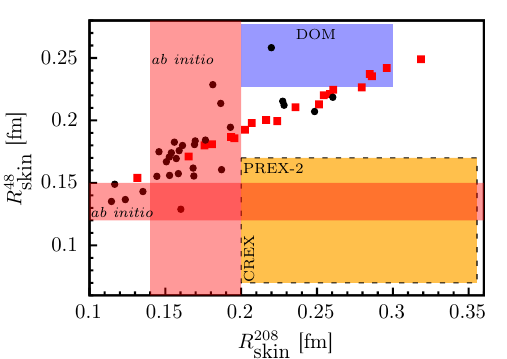}
      }
   \end{minipage}
   \caption{ \label{fig:skinc}
    The dashed rectangle represents the CREX and PREX-2 analysis~\cite{CREX22,PREX21}. 
The shaded rectangle labeled DOM represents the DOM results for ${}^{208}$Pb and $^{48}$Ca~\cite{Mahzoon:2017,Atkinson:2020}. Smaller squares and circles refer to relativistic and nonrelativistic mean-field calculations, respectively, cited in Ref.~\cite{Horowitz14}.
The \textit{ab initio} predictions from Ref.~\cite{Hagen:2016} for $^{48}$Ca and Refs.~\cite{Hu:2022,Hu:2024} for $^{208}$Pb are represented by horizontal and vertical bands labeled \textit{ab initio}, respectively.  All uncertainties are reported at the $1\sigma$ level. Figure adapted from Refs.~\cite{Horowitz14,Atkinson:2020}.
   }
  
\end{figure}

As demonstrated in the previous section, our constrained self-energies for $^{48}$Ca and $^{208}$Pb utilize both scattering and bound-state data for a robust picture of nuclei. These fits resulted in thick skins in both ${}^{48}$Ca, $R^{\textrm{DOM}48}_\textrm{skin}=0.25 \pm 0.023$ fm, and ${}^{208}$Pb, $R^{\textrm{DOM}208}_\textrm{skin}=0.25 \pm 0.05$ fm using the uncertainty quantification clarified in Refs.~\cite{Mahzoon:2017,Atkinson:2020}. 
These results are represented by the shaded box labeled DOM in Fig.~\ref{fig:skinc} which is north of the overlapping regions of CREX and PREX-2 (see dashed rectangle). 
Also included in Fig~\ref{fig:skinc} is the coupled-cluster result for $^{48}$Ca from Ref.~\cite{Hagen:2016} as a horizontal band, the \textit{ab initio} results for $^{208}$Pb reported in Refs.~\cite{Hu:2022,Hu:2024} as a vertical band, and both relativistic and nonrelativistic mean-field calculations represented by squares and circles, respectively~\cite{Horowitz14}.
Relativistic and nonrelativistic mean-field calculations cited in Ref.~\cite{Horowitz14} are represented by squares and circles, respectively.

\begin{figure}[t]
   \begin{center}
   \includegraphics[width=0.48\linewidth]{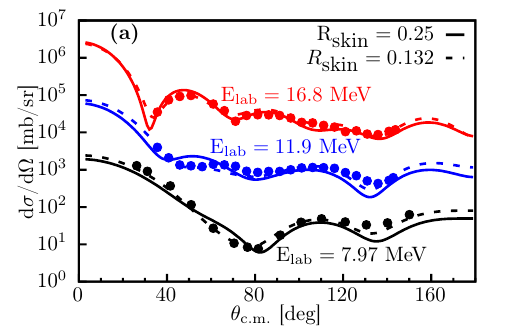}
   \includegraphics[width=0.48\linewidth]{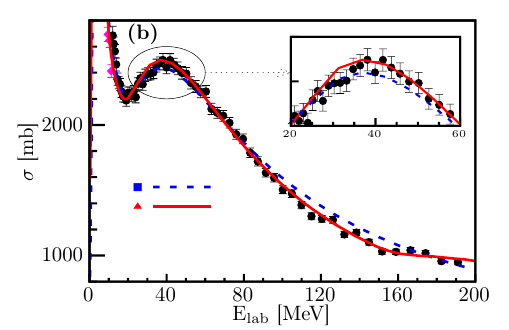}
\end{center}
   \begin{center}
   \includegraphics[width=0.48\linewidth]{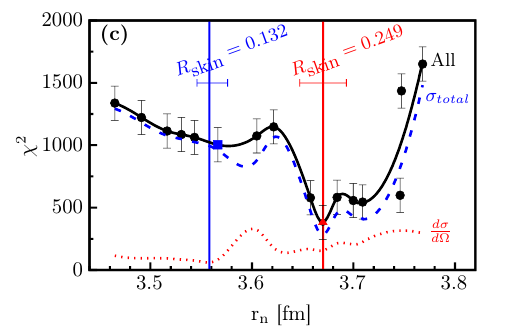}
\end{center}
\caption{\label{fig:skin_landscape}
(a) Comparison of experimental $n$+$^{48}$Ca elastic-scattering angular distributions \cite{Hicks:1990,Mueller:2011} to the best DOM fit of all data (solid curves) and to a constrained fit with the skin thickness forced to $R_\textrm{skin}$=0.132~fm (dashed curves) consistent with the \textit{ab initio} and CREX values. (b) Comparison of the experimental total neutron cross sections of $^{48}$Ca (diamonds \cite{Harvey:1985}, circles \cite{Shane:2010}) to DOM fits with constrained values of $R_n$. The curve labeled with a triangle is for the $R_n$ value of our best fit, while the curve labeled with a square is for a value consistent with \textit{ab initio} and CREX values (see panel (c)). (c) The $\chi^2$ from fitting all data (solid curve) and its contribution from fitting the elastic-scattering angular distributions and total neutron cross section (short-dashed and long-dashed curves respectively). Each point corresponds to a fit around its value of $R_n$. Figure adapted from Ref.~\cite{Mahzoon:2017}.} 
\end{figure}

At the time of our calculations, CREX had not been reported and only the first PREX experiment with large uncertainty had been reported, meaning that there was not an easy metric to gauge the accuracy of our predictions. Therefore, we took advantage of the unique characteristic of the DOM to explore which measurements, in either the bound or scattering domains, provide signatures of the neutron skin.
To accomplish this, additional $^{48}$Ca fits were performed in which selected values of $R_n$ are forced (i.e. heavily weighted) in the corresponding $\chi^2$ minimization~\cite{Dickhoff:2017}. 
This is achieved by varying the radius parameters of the main real potential
($r^{HF}_{n}$ and $r^{HFasy}_{n}$~\cite{Mahzoon:2017}) and refitting the other asymmetry-dependent parameters.  The weighted $\chi^2$ as a function 
of the calculated $R_n$ is plotted as the points (traced by the solid black line) in Fig.~\ref{fig:skin_landscape}(c) and the 
absolute minimum at $R_n$=3.67~fm corresponds to the skin thickness of $R_\textrm{skin} =0.25$~fm.
There is some fine-scale jitter in the variation of $\chi^2$ with $R_n$. 
To concentrate on the larger-scale variation, the data points shown in Fig.~\ref{fig:skin_landscape}(c) are local averages with the error bars giving the range of the jitter.    

The location of the \textit{ab initio} coupled-cluster result~\cite{Hagen:2016} is also indicated at $R_n \sim $3.56~fm as a blue square. The shown $\chi^2$ has been subdivided into its contributions from its two most important
components (dashed curves); from the elastic-scattering angular distributions and from the total neutron cross sections. The former has a smaller sensitivity to $R_n$, and its $\chi^2$ is slightly lower for
the smaller values of $R_n$ which are more consistent with the \textit{ab initio} and CREX results as illustrated in Fig.~\ref{fig:skin_landscape}(a) where a fit with a forced value of $R_\textrm{skin}$=0.132 is compared to the
best DOM fit and to the data. While this alternative calculation improves the 
reproduction of these data, the deviations of both curves from the data are typical of what one sees in global optical-model fits. 
In addition, the available experimental angular distributions only cover a small range of bombarding energies (7.97 to 16.8 MeV) and may not be typical of other energies. 

The total cross section exhibits larger sensitivity and the experimental data cover a large range of neutron energies (6 to 200~MeV). Two data sets are available (circles and diamonds) but are inconsistent by
$\sim$10\% at $E_{lab}\sim$10 MeV, where their ranges overlap. The high-energy data set~\cite{Shane:2010} (circles) was used in the DOM fit as it was obtained with $^{48}$Ca metal, while the low-energy
set~\cite{Harvey:1985} (diamonds) employed $^{48}$CaCO$_{3}$ and required a subtraction of $\sim$70\% of the signal due to neutron absorption from the CO$_{3}$ component. Therefore, the $\chi^2$ contribution
is displayed only from the high-energy set.  This $\chi^2$ exhibits a broad minimum from $R_n$= 3.66 to 3.75~fm allowing values of $R_\textrm{skin}$ up to 0.33~fm.  

It appeared that the total cross section provided a strong constraint on the neutron skin (as an example of a scattering observable that can affect bound-state observables through the dispersion relation). Faced with the thin skin reported by CREX, it appears that we did not attribute enough uncertainty in the total cross-section data to allow a wider range of skin values. This concept will be explored in future DOM investigations of $R^{48}_\textrm{skin}$ in which the CREX measurement is included in the fit. It is possible that increasing the uncertainty in the high-energy $\sigma_\textrm{tot}$ data would allow for skin values consistent with CREX (i.e. the blue square in Fig.~\ref{fig:skin_landscape}(c)) to have $\chi^2$ values comparable to those of the current DOM fit. Furthermore, similar to the analysis that resulted in Fig.~\ref{fig:skin_landscape}, it will be interesting to see how the CREX constraint alters other aspects of the DOM self-energy, even non-observables features such as the shape of the spectral functions (see Fig.~\ref{fig:spectral_n}) and the momentum distributions (see Fig.~\ref{fig:kdist}). 

Provided with a sufficiently-complete set of data, which is the case for protons in $^{48}$Ca, the DOM framework allows for accurate predictions (see Sec.~\ref{sec:dom-predict}). 
The thin skin of CREX demonstrates that, unlike protons, there are not sufficient experimental data for neutrons in $^{48}$Ca to accurately predict the neutron skin. The number of proton elastic-scattering data sets at different energies shown for $^{208}$Pb in Fig.~\ref{fig:elastic_pb208} is representative of $p+^{48}$Ca, while the three data sets in Fig.~\ref{fig:skin_landscape}(a) displays all available data for $n+^{48}$Ca elastic-scattering. Furthermore, there are only neutron total cross-section data, and no reaction cross-section data exists at any energy in $^{48}$Ca. Thus, even at positive energies, the DOM neutrons are not constrained nearly as well as protons. 
With more neutron scattering data in $^{48}$Ca, the DOM could provide a better prediction of $R_\textrm{skin}^{48}$. Furthermore, the inclusion of the CREX data point will provide a much needed constraint below the Fermi energy, bringing the neutron data set closer to "completeness" (in the sense of constraining the DOM).

To accommodate the thin skin extracted by CREX, one would expect the distribution of neutrons to favor a configuration with more neutrons in the interior of $^{48}$Ca. This concentration of neutrons near the origin implies an increase in the fraction of high-momentum neutrons thanks to the Heisenberg uncertainty principle. This could lead to a larger percentage of high-momentum neutrons than protons, which would be a departure from the current DOM picture (see Fig.~\ref{fig:kdist}) as well as from the evidence suggested by the CLAS experiments on other asymmetric nuclei. Currently this is speculation, but we are exploring new DOM fits using CREX as an additional constraint so we can reach a better understanding. It could turn out that the size of $^{48}$Ca is inadequate to apply bulk nuclear properties to. We observed this in Ref.~\cite{Atkinson2020B} where we consider the interiors of $^{48}$Ca and $^{208}$Pb as representing saturated nuclear matter. We found that the smaller size of $^{48}$Ca compared to $^{208}$Pb is harder to connect with saturated nuclear matter. 

  



\begin{figure}[b]
   \begin{minipage}{\linewidth}
      \makebox[\linewidth]{
         \includegraphics[scale=1.0]{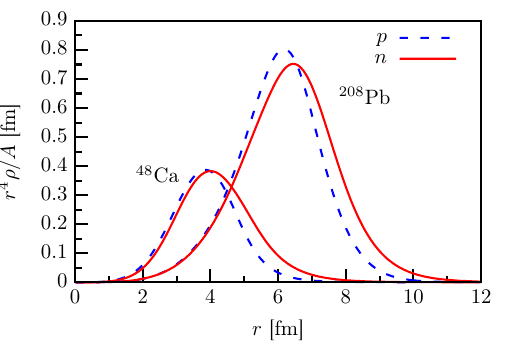}
      }
   \end{minipage}
   \caption[Neutron and proton point distributions in $^{208}$Pb weighted by $r^4$]{\label{fig:dist_comp}Neutron (red solid line)
      and proton (blue dashed line) point distributions in $^{208}$Pb and $^{48}$Ca weighted by $r^4$ while
      normalized to particle number. Figure adapted from Ref.~\cite{Atkinson:2020}.
   }
\end{figure} 

The neutron and proton point distributions in $^{208}$Pb and $^{48}$Ca, weighted by $r^4$ and normalized by particle number, are shown in Fig.~\ref{fig:dist_comp}. 
The difference between proton and neutron distributions is highlighted by the $r^4$ factor which is employed when integrating the particle distributions to calculate the RMS radii.
The DOM predictions of the neutron skin of $^{40}$Ca, $^{48}$Ca, and $^{208}$Pb are shown in Table.~\ref{table:skins_normalized}, where it is evident that the DOM neutron skins of $^{48}$Ca and $^{208}$Pb are very
similar. Since $^{208}$Pb and $^{48}$Ca have similar asymmetry parameters, indicated by $\alpha_\text{asy} = (A-Z)/A$ in Table~\ref{table:skins_normalized}, it may seem reasonable that they have similar neutron
skins. 
However, the particle distributions of $^{208}$Pb and $^{48}$Ca in Fig.~\ref{fig:dist_comp}, even though normalized by particle number, 
are quite distinct due to the size difference of the nuclei. In light of this, the neutron skin of $^{208}$Pb is biased to be larger by the increase in the RMS radii of the proton and
neutron distributions. Thus, an interesting comparison can be made by normalizing
$R_\textrm{skin}$ by $R_p$, 
\begin{equation}
   \tilde{R}_\textrm{skin} = \frac{1}{R_p}R_\textrm{skin} = \frac{R_n}{R_p}-1,
   \label{eq:skin_normalized}
\end{equation}
where $\tilde{R}_\textrm{skin}$ is the normalized neutron skin thickness.  This normalization serves to remove size dependence when comparing neutron skins of different nuclei.  The result of this normalization
is shown in Table~\ref{table:skins_normalized}. The difference between the normalized skins of $^{208}$Pb and $^{48}$Ca in Table~\ref{table:skins_normalized} reveals that the RMS radius of the neutron
distribution does not simply scale by the size of the nucleus for nuclei with similar asymmetries. While it is true that the nuclear charge radius scales roughly by $A^{1/3}$ (and by extension so does $R_p$), the
same cannot be said about $R_n$. 

If one is to scale by the size of the nucleus, then the extension of the proton distribution due to Coulomb repulsion (which scales with the number of protons) should also be considered. Since $^{208}$Pb has
four times as many protons as $^{48}$Ca, the effect of Coulomb repulsion on the neutron skin of $^{208}$Pb could be up to four times more than its effect on the $^{48}$Ca neutron skin, which can reasonably
be taken from the predicted skin of $-0.06$~fm in $^{40}$Ca. In order to further investigate the effects of the Coulomb force on the neutron skin, we removed the Coulomb potential from the DOM self-energy.  In
doing this, the quasihole energy levels become much more bound, which increases the number of protons. To account for this, we shifted $\varepsilon_F$ such that it remains between the particle-hole gap of the
protons in $^{208}$Pb, corresponding to a shift of 19~MeV. Removing the effects of the Coulomb potential leads to an increased neutron skin of 0.38~fm. The results of the normalized neutron skins with Coulomb
removed are listed in Table~\ref{table:skins_normalized} for each nucleus, where it is clear that the Coulomb potential has a strong effect on the neutron skin. This points to the fact that the formation of a
neutron skin cannot be explained by the asymmetry alone. Whereas the asymmetry in $^{48}$Ca is primarily caused by the additional neutrons in the f$7/2$ shell, there are several different
additional shell fillings between the neutrons and protons in $^{208}$Pb. 
It is evident that these shell effects make it more difficult to predict the formation of the neutron skin based on macroscopic properties alone.

\begin{table}[h]
   \caption[DOM Predicted neutron skins for $^{40}$Ca, $^{48}$Ca, and
   $^{208}$Pb.]
   {\label{table:skins_normalized}DOM Predicted neutron skins for $^{40}$Ca, $^{48}$Ca, and
   $^{208}$Pb. Also shown are the neutron skins normalized by $R_p$, denoted as $\tilde{R}_\textrm{skin}$, as well as neutron skins with the Coulomb potential removed from the self-energy, denoted as $R^{noC}_\textrm{skin}$. The last entry is the normalized neutron skin with Coulomb removed, $\tilde{R}^{noC}_\textrm{skin}$.}
   
   \begin{center}
      {\renewcommand{\arraystretch}{1.35}
      \begin{tabular}{cccc}
         \hline
         \hline
         Nucleus  &  $^{40}$Ca & $^{48}$Ca & $^{208}$Pb \\
         \hline
         $\alpha_\text{asy}$ & 0 & 0.167   & 0.211  \\
         \hline
         $R_p$ & $3.39$~fm &  $3.38$~fm  &  $5.45$~fm \\
         \hline
         $R_n$ & $3.33$~fm  &  $3.63 \pm 0.023$~fm  &  $5.70 \pm 0.05$~fm \\
         \hline
         $R_\textrm{skin}$   & $-0.06$~fm  & $0.25\pm0.023$~fm & $0.25\pm0.05$~fm \\
         \hline
         $\tilde{R}_\textrm{skin}$ & $-0.017$ & $0.070\pm0.0067$ & $0.046\pm0.0092$ \\
         \hline
         $R^{noC}_\textrm{skin}$   & $0$~fm  & $0.309\pm0.023$~fm & $0.380\pm0.05$~fm \\
         \hline
         $\tilde{R}^{noC}_\textrm{skin}$ & $0$ & $0.089\pm0.0067$ & $0.070\pm0.0092$ \\
         \hline
         \hline
      \end{tabular}
   }
\end{center}
\end{table}

\section{Conclusions}
\label{sec:conclusions}

We have reviewed a nonlocal dispersive optical-model analysis of ${}^{48}$Ca and $^{208}$Pb in which we fit elastic-scattering angular distributions, absorption and total cross sections, single-particle energies, charge densities, ground-state binding energies, and particle numbers. When sufficient data is available to constrain our self-energies, the DOM is capable of accurate predictions. With our well-constrained self-energies we report non-negligible high-momentum content in both $^{48}$Ca and $^{208}$Pb, which is consistent with the experimental observations at JLAB~\cite{CLAS:2006,Duer:2018,Hen:2017}.
Spectroscopic factors are automatically generated and reproduce $^{48}$Ca$(e,e'p)^{47}$K experimental momentum distributions and those predicted in $^{208}$Pb appear consistent with the most up-to-date analysis of the $(e,e'p)$ reaction for the last valence proton orbit~\cite{Ingo91}.
Furthermore, these spectroscopic factors explain the reduction of the form factors of high spin states obtained in inelastic electron scattering~\cite{Lichtenstadt79} lending support to the interpretation of Ref.~\cite{Vijay84}.
The thick skin predicted in $^{208}$Pb ($R^{208}_\textrm{skin} = 0.25\pm 0.05$) is in agreement with PREX-2 while that predicted in $^{48}$Ca
($R^{48}_\textrm{skin} = 0.25\pm 0.023$) is not consistent with CREX.  With more neutron scattering data in $^{48}$Ca, the DOM could provide a better prediction of
$R_\textrm{skin}^{48}$. Including the CREX result in a DOM fit of $^{48}$Ca would provide a much needed constraint, bringing the neutron data set closer to ``completeness".

To reproduce the reduced neutron RMS radius reported by CREX, we expect that the neutron distribution in $^{48}$Ca would shrink such that more neutrons get concentrated in the interior of $^{48}$Ca. This redistribution would translate to increased high-momentum neutrons which could invert the hierarchy of the current DOM fit in which there is a higher percentage of high-momentum protons than neutrons (see Fig.~\ref{fig:kdist}), counter to the evidence suggested by the CLAS experiments on other asymmetric nuclei~\cite{CLAS:2006,Duer:2018}. Currently this is speculation, but we are exploring new DOM fits using CREX as an additional constraint so we can reach a better understanding. We must also consider the possibility that the size of $^{48}$Ca is inadequate for extracting/applying bulk nuclear properties. The shell-closure of the $f7/2$ neutrons in $^{48}$Ca, for example, could be playing a stronger role in the formation of the skin than the EOS. Similarly, it is possible that this $np$ dominance picture is distorted by finite-nucleus effects that are not negligible in $^{48}$Ca. The DOM provides a unique perspective of the nucleus in that we can link these entirely different measurements through the dispersion relation in order to reach a deeper understanding of the relation between the EOS (and hence exotic objects such as neutron stars) and finite nuclei. 



The DOM analysis provides an alternative approach to the
multitude of mean-field calculations that provide a large variety of results
for the neutron skins of ${}^{48}$Ca and ${}^{208}$Pb~\cite{Horowitz14} while also
contrasting with the \textit{ab initio} result of Ref.~\cite{Hagen:2016} for ${}^{48}$Ca and Refs.~\cite{Hu:2022,Hu:2024} for ${}^{208}$Pb.
The experiments employing parity-violating elastic electron scattering on
these nuclei~\cite{crex:2022,prex2:2021} therefore remain the most unambiguous approach
to determine the neutron skin.
A systematic study of more nuclei with similar asymmetry, $\alpha_\text{asy}$,
to $^{208}$Pb and $^{48}$Ca would help in determining the details of the
formation of the neutron skin. This will lead to a better understanding of the
EOS, which is
vital in the current multi-messenger era onset by the first direct
detection of a neutron star merger~\cite{LIGO:2017}.

\section*{Conflict of Interest Statement}

The authors declare that the research was conducted in the absence of any commercial or financial relationships that could be construed as a potential conflict of interest.

\section*{Author Contributions}

The authors provided approximately equivalent contributions to the conceptual and practical aspects of this review.

\section*{Funding}
This work was performed under the auspices of the U.S. Department of Energy by Lawrence Livermore National Laboratory under Contract DE-AC52-07NA27344 and was supported by the LLNL-LDRD Program under Project No. 24-LW-062. This work was also supported by the U.S. National Science Foundation under grants PHY-1912643 and PHY-2207756.

\section*{Acknowledgments}
The authors acknowledge the important early contributions to this research from Bob Charity, Hossein Mahzoon, Cole Pruitt, and Lee Sobotka.

\bibliographystyle{Frontiers-Vancouver} 
\bibliography{skin-review}

\end{document}